\documentclass[12pt]{iopart}
\usepackage{graphicx}
\usepackage{cite}
\usepackage{bm}
\begin{document}
\title[Magneto-transport of graphene and QPTs]{Magneto-transport of graphene and quantum phase transitions in the quantum Hall regime}

\author{Mario Amado$^{1,2}$\footnote{Present address: SNS-NEST \& CNR, Piazza San Silvestro 12, 56127 Pisa, Italy}, Enrique Diez$^{1}$, Francesco Rossella$^{3}$, Vittorio Bellani$^{3}$, David L\'opez-Romero$^{4}$ and Duncan K Maude$^{5}$}

\address{
$^{1}$ Laboratorio de Bajas Temperaturas, Universidad de Salamanca, 37008 Salamanca, Spain \\
$^{2}$ GISC-QNS, Dpto de F\'{\i}sica de Materiales, Universidad Complutense, 28040 Madrid, Spain \\
$^{3}$ Dipartimento di Fisica ``A. Volta'' and CNISM, Universit\`{a} degli Studi di Pavia, 27100 Pavia, Italy \\
$^{4}$ CT-ISOM, Universidad Polit\'ecnica de Madrid, 28040 Madrid, Spain \\
$^{5}$ Laboratoire National des Champs Magn\'etiques Intenses, 38042 Grenoble, France
}

\ead{mario.amado@nano.cnr.it}

\begin{abstract}
We studied the magneto-transport in SiO$_2$ substrate-supported monolayer graphene and the quantum phase transitions that characterize the quantum Hall regime, using magnetic fields up to 28 T and temperatures down to $4$ K. The analysis of the temperature dependence of the Hall and longitudinal resistivity reveals new non-universalities of the critical exponents of the plateau-insulator transition. These exponent depends on the type of disorder that governs the electrical transport, which knowledge is important for the design and fabrication of new graphene nano-devices.
\end{abstract}

\maketitle


\section{Introduction}
The discovery of the quantum Hall effect (QHE) in two dimensional electron gas (2DEG) opened a new pathway in the study of quantum phase transitions (QPTs). In the quantum Hall regime the Hall resistivity ($\rho_{xy}$), measured as a function of the magnetic field or of the charge density, exhibits plateaus while the longitudinal resistivity ($\rho_{xx}$) vanishes, as a consequence of the 2DEG quantum Hall localization. In the region between two adjacent plateaus, in correspondence to the 2DEG delocalization, $\rho_{xx}$ shows Shubnikov-de Haas (SdH) peaks which has a temperature dependent width. In this region a QPT takes place, called plateau-plateau (PP) transition, which has been studied since the pioneering experimental works by Tsui and coworkers~\cite{weiprb,tsui} and the theoretical page-to-page one by Pruisken $et al.$~\cite{pruisken-pp} In the limit of zero temperature the concept of delocalization can be understood assuming that the localization length $\xi$ diverges for some ranges of the Fermi energy. Then, there exist a singular energy $E_c$ in the energy spectrum where $\xi$ diverges with a power-like law $\xi\sim(E-E_c)^{-\gamma}$, being $\gamma$ its critical index that has been theoretically calculated and supposed to be universal. The first works dealing with the calculation of this index neglected the electron-electron (\textit{e-e}) interaction and revealed a constant value $\gamma = 2.38$~\cite{CC,Lee,Cho} for the localization length critical index. More recent and accurate calculations, carried out taking into account the \textit{e-e} interaction, give an higher value $\gamma\simeq2.61(1)$~\cite{Slevin,Amado-CC}. Indications emphasizing the importance of interactions for the localization phenomenon also come from recent experiments on high-mobility low-density 2D electron structures,
which have given evidence for the existence of an interaction-mediated metal-insulator transition ~\cite{Anissimova}. Experimental measurements have also
lately revealed that electronic localization in graphene, in the quantum Hall regime, is not entirely
dominated by single-particle physics, but rather a competition between the underlying disorder
and the repulsive Coulomb interaction exists ~\cite{Martin}. Indeed, the effect of interactions near the QPTs, and in particular the role of multifractality, are not well understood at the moment. For instance, at the Integer Quantum Hall transition shortrange interactions seem to be irrelevant, in an renormalization group sense, at the critical point ~\cite{Lee2}, (i.e. the critical exponent $\gamma$ for the localization length and the multifractal spectrum remain the same as in the non-interacting problem). The $1/r$ Coulomb interaction is relevant, however, and should drive
the system to a novel critical point ~\cite{Baranov}. Consequently, how the value of the critical exponent $\gamma$
is affected, is far from being solved.

Experimentally it is possible to obtain $\gamma$ from the magneto-transport measurement using an indirect method which relies on the calculation of the critical exponent $\kappa$. This method exploits the relation $\kappa = p/2\gamma$, being $p$ the temperature exponent of the inelastic scattering length ($L_\Phi \propto T^{-p/2}$), widely accepted~\cite{Huckestein} to be equal to $2$ within the framework of the Anderson model (when the disorder is of the alloy type an there is no electron-electron interaction). The pioneering experimental work of Wei \emph{et al.}~\cite{tsui} gave $\kappa = 0.42(4)$, a value that has been recently reproduced extending the study to temperatures down to the milli-Kelvin regime~\cite{Wanli}. As explained above $\kappa$ was accepted to be universal for PP transitions under the Anderson model but recent experiments have shown that the value of the exponent depends on the type of disorder. Indeed Wanli Li \emph{et al.}~\cite{Wanlialloy} observed different values for $\kappa$ in AlGaAs/InGaAs heterostructures by changing the density of Al, so that it could be possible that the clustering of Al implies a change in the regime of the transition. This would mean that the Anderson model is no more applicable in systems with long range disorder. This hypothesis has been strengthened by experimental studies that provided $\kappa$ values between 0.2 and 0.9~\cite{wakabayashipp,koch}. Indeed in presence of this type of disorder or when interactions are playing an important role, the electron transport is better described by a percolation picture through the so-called saddle points rather than by the Anderson localization so that the scaling regime of the QPTs has still to be reached and the obtained non-universal value at most represents a cross-over behavior.

The magneto-transport measurements in a perfect two dimensional electron systems, such as graphene, in the quantum Hall regime, presents another QPT, referred as plateau-insulator (PI) transition, which is the transition from the last Hall resistivity plateau of the integer quantum
Hall effect (IQHE) ($\nu = 1$ in semiconductor $2$DEGs and $\nu = \pm2$ in the case of graphene) to an insulating phase which persists at higher magnetic fields. Magneto-transport experiments in a semiconductor 2DEG have shown that $\rho_{xx}$ passes from a metallic phase (in the delocalized state or SdH peak) which value decreases by decreasing $T$ , to an insulating phase where $\rho_{xx}$ increases by decreasing $T$. In such transition there exists a T-independent crossing point corresponding to a critical magnetic field $B_c$ where all the isotherms collapse. The value obtained for the critical exponent of this transition in semiconductor samples (InGaAs/InP and InGaAs/GaAs) ~\cite{Schaijk,deLang,ponomarenko1,Hilke1,Pruisken,ponomarenkoTH} was $\kappa = 0.58$, but there still exist some controversy whether this transition belongs to the same universality class of the PP one.

The recent discovery of graphene has implied a revision of the 2DEG knowledge since it is the first truly two dimensional system studied and its properties differ markedly from the ones of the semiconductor-based 2DEG. Indeed, magneto-transport experiments performed in graphene have shown a new type of the IQHE~\cite{novoselov2005,stormer2005}, as a consequence of the fourfold degeneracy of the charge carriers~\cite{guinea2}, and the half-filling of the $n = 0$ Landau level (LL)~\cite{gusynin1} . The fourfold degeneracy of the graphene carriers arises from the usual twofold spin degeneracy and from a novel twofold valley one. These two properties led to assume a special sequence of filling factor in graphene $\nu = \pm2,\pm6,\ldots$, in contrast with the well-established sequence of filling factors typical of the IQHE in semiconductor 2DEG ($\nu = 1,2,\ldots$). This sequence of filling factors is no longer kept in clean graphene samples, and new integer plateaus close to the charge neutrality point (CNP) at $\nu = 0,\pm1,\pm3,\pm4$ appear, as well as fractional ones~\cite{Zhang,Abanin,Jiang,Ong1,Giesbers,Du}. These exotic states have been observed in the presence magnetic field strong enough to lift the spin and valley degeneracy implying symmetry breaking effects and depend strongly on the homogeneity and purity of the graphene sample.

SiO$_2$ substrate-supported monolayer graphene samples are usually far to be clean or homogeneous in sharp contrast with suspended or BN supported ones. Some indicators of their quality are a high value of the mobility ($>20000 cm^2V^{-1}s^{-1}$), the proximity of the charge neutrality point (CNP) to $0$~V, the narrowness of the Dirac peak and its increase when applying an external magnetic field. Moderate mobility samples (with values in the range of $10^3 cm^2V^{-1} s^{-1}$) and a CNP far from $0$~V indicate the presence of noticeable disorder like charged impurities, substrate effects, corrugations and strain (see the work of Mucciolo and coworkers for a complete review~\cite{Mucciolo}). Since it is not easy to control the homogeneity of the sample, it still remains unclear which type of disorder or interaction (charged impurities, alloy, etc.) is the dominant one in each sample. For this major reason the study of the effect of the disorder and interactions in the QHE in graphene is at a preliminary stage yet. Previous experiments have shown that in the vicinity of the the CNP $\rho_{xx}$ may either decrease~\cite{Abanin}, increase~\cite{Ong1} at $\nu \sim 0$ with decreasing temperature or exhibit a metallic state~\cite{Wiedmann} without traces of an insulating one when crossing the CNP resulting of the coexistence of electron and hole puddles~\cite{Wiedmann,Poumirol}. These observations have fueled the debate on the existence and origin of an insulating phase close or at the CNP at high magnetic fields. This confusion is partially clarified in the work by Zhang \emph{et al.}~\cite{Zaliznyak} where it is clearly established that there are two different insulating regimes on the N=$0$ LL in SiO$_2$ substrate-supported monolayer graphene samples (in the vicinity of not at the CNP) and that they appear accordingly to the quality of the sample. A first transition to a quantum Hall insulator occurs due to existence of disorder in the sample and can be followed by a second one to a bulk collective insulator state (a pinned Wigner crystal has been suggested) at half LL filling regime. This latter transition arises from the symmetry breaking and would be observable in samples that exhibits high mobility ($>20000 cm^2V^{-1}s^{-1}$), then it would be hindered by the presence of the disorder in low mobility samples.

As explained above, the non-interacting electrons model does not fully describe the critical phenomena of the IQHE so that the model must be improved, in order to clarify the mechanism that control the interplay between disorder and $e-e$ interaction on the electron transport in graphene in the quantum Hall regime. In this work we try to shed more light to the case. Actually, in a previous work~\cite{Amado-PI} we observed the $\nu = 0$ plateau in a graphene sample and we reported preliminary measurements on the PI QPT obtaining the value $\kappa = 0.58(3)$, in agreement with the one obtained with semiconductor alloys but in a limited range of temperatures. In the present work we present a detailed study of the quantum Hall magneto-transport in various regimes of carrier concentration (which were tuned by means of a gate voltage $V_G$ applied to the back-gate of the graphene Hall bar) in a moderate Drude mobility sample. We have carried out the measurements from liquid He temperature up to 230 K. In this way we explore the interplay between the Coulomb interaction, the thermal energy and the disorder landscape probed by the charge carriers and the overall effect on the transport in the quantum regime. In particular we reveal the non universality of the critical exponents when the spin-valley degeneracy is not lifted, a fact which give new insights in the knowledge of standard quantum Hall insulator $\nu = 0$ state.

\section{Materials and Methods}

In this work we have used a single graphene monolayer sample used that was obtained by mechanical exfoliation of natural graphite and deposited onto a Si wafer with a thermally grown $300$ ~ nm SiO$_2$ top layer. The ohmic contacts of sample were patterned by means of nano-lithography using PMMA resist whereas the graphene was not patterned into Hall bar geometry by means of plasma etch in order to reduce the role of defects from edge states and the doping introduced by the mesa-etching process. The deposition of the ohmic contacts (an adhesion layer of $5\,$nm of Ti followed by $5\,$nm of Au) was performed in a ultra ($10^{-7}$ mbar) high vacuum electron-beam evaporator. The aspect ratio of the sample W/L = Width/Length was $5.1$(nm)/$10.6$(nm)$\simeq 0.48$. Magneto-transport measurements were carried out by a $4$-probe low frequency AC lock-in method. We used a fixed frequency of the lock-in of $10.66$~Hz and a fixed excitation current of $10$ nA applying a drop of voltage of $1$~V on a $100$~M$\Omega$ resistance. A source-meter Keithley $2400$ grounded with a $1$ M$\Omega$ in-series resistance and current leak lower than $2$nA at every voltage was used as back-gate voltage source. The temperature was measured with a calibrated $4$-pins RuO$_2$ resistance. The sample was placed into a $^4$He cryostat with a variable temperature insert that led us to vary the temperature in the range from $~4$~K up to room temperature. A resistive $20$~MW magnet allowed us to obtain magnetic fields up to $28$~T. The acquisition rate during the resistivity measurements was $3.2$ points per second, and the rate of variation of the magnetic field was of the order of $60-120$ Gauss per second. We recorded the signal decreasing $B$ from 28 T value to zero in order to exploit higher thermal stability, and we monitored the out-of-phase component of the signal to verify that it did not affect $\rho_{xx}$ at high $B$.

\begin{figure}[ht]
\centerline{\includegraphics[width=90mm,clip]{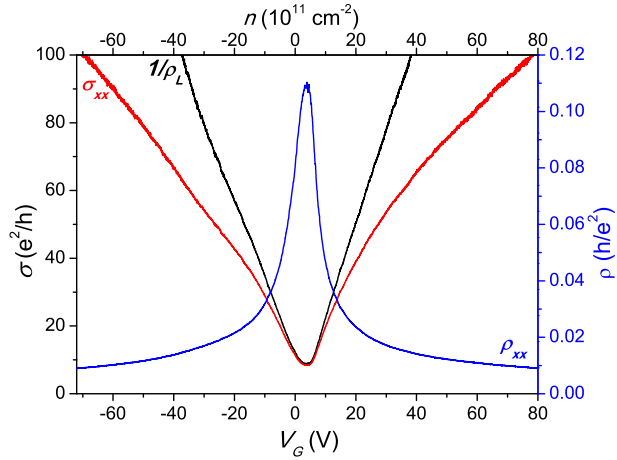}}
\caption{Longitudinal conductivity a function of both the density of carriers and the gate voltage calculated as $\sigma_{xx} = \rho_{xx}/(\rho^2_{xx}+\rho^2_{xy})$, measured at $4.2\,$K and $B = 0$. The carrier density $n$ at each sides of the Dirac point was obtained with an approach similar that the one used in Ref.~\cite{Amado-PI}}\label{density-vs-VG}
\end{figure}

Fig.~\ref{density-vs-VG} reports the longitudinal resistivity $\rho_{xx}$ (blue solid curve) measured at $B = 0$ and $T = 4.2$ K as a function of the applied gate voltage and shows that the CNP appears at V$_G=3.89\,$V. The charge density $n$ was calculated using an approach similar to the one found in our previous work~\cite{Amado-PI}. The measurements of the conductivity at a given $V_G$ at low magnetic fields and at different gate voltages allowed us to extract $n$ both in the electron-like and hole-like regimes. The bare longitudinal conductivity $sigma_{xx}$ shows a noticeable sublinear behavior as a function of V$_G$. Following a previous work by Morozov \emph{et al.}~\cite{Morozov} we have extracted a constant resistivity $\rho_s\simeq150\Omega$ becoming then the longitudinal conductivity $1/\rho_L=1/[\rho(V_G)-\rho_s]$ perfectly linear for the whole range of V$_G$ thus we can then affirm that within our sample macroscopic inhomogeneities are negligible. As shown in the work by Morozov \emph{et al.}~\cite{Morozov} there are two different contributions for the resistivity in doped graphene. The one for short-range scatterers $\rho_s$ that is n-independent and a second $\rho_L\propto1/n$ for long-range ones. Therefore at high carrier densities the role of long-range disorder will be  much less effective and we can expect the experimental critical exponents closer to the obtained for the ideal short-range Anderson disorder.

\section{Experimental derivation of critical exponents}

The most common way to obtain the value of the critical exponent for the PP transition was postulated by Wei \emph{et al.}~\cite{tsui} translating, by a finite-size scaling method, the power-law dependence of the localization length into a temperature scaling form. The new scaling law rules the maximum value of the derivative $\partial \rho_{xy}/\partial B$ and/or the distance $\Delta B$ in magnetic field between the two extreme in $(\partial \rho_{xx}/\partial B)^{-1}$. The latter derivative is also proportional to the width of the SdH peaks~\cite{Wanli} so that the critical exponent of the PP transition can be obtained by means of the two relations

\begin{equation}\label{power-law-K-B}
        \Delta B \propto T^\kappa\ ,
\end{equation}
   \begin{equation}\label{power-law-K-rho}
        \left( \frac{\partial \rho_{xy}}{\partial B}\right)_{max} \propto T^{-\kappa}\ ,
\end{equation}
being $\kappa$ the critical exponent that can be therefore extracted from the measurement of $\rho_{xy}$ as a function of $B$.

Giesbers \emph{et al.}~\cite{ppgiesbers} gave the first value for the critical exponent of the PP transition in graphene measuring the evolution of the maximum slope of $\partial \sigma_{xy}$/$\partial \nu$ and the inverse of the width of the derivative of the SdH peaks at different temperatures, using the as driving parameter the voltage $V_G$. They obtained $\kappa = 0.41(4)$ for the first and second LLs using a equation equivalent to \ref{power-law-K-rho} and $\kappa = 0.37(5)$ for the first LL using an equation equivalent to \ref{power-law-K-B}.

As regards the other QPT investigated in this work, namely the MI transition, the critical exponent can be obtained using the empirical law:~\cite{Shahar1,Pruisken,Amado-PI}

\begin{equation}
\label{scaling1} \rho_{xx} = e^{-\Delta \nu/\nu_0(T)}\,,
\end{equation}
with $\Delta \nu = 1/B - 1/B_c$ ($\nu$ here should not be confused with the Landau level filling factor) and $B_c$ the temperature independent critical field. The experimental $\rho_{xx}$ measured at different temperatures are fitted using \ref{scaling1} in order to obtain the associated critical exponent
\begin{equation}
\label{scaling-PI} \nu_{0}\propto T^{\kappa}.\
\end{equation}

\section{Results and discussion: Plateau-Insulator transition}

\begin{figure}[ht]
\centerline{\includegraphics[width=90mm,clip]{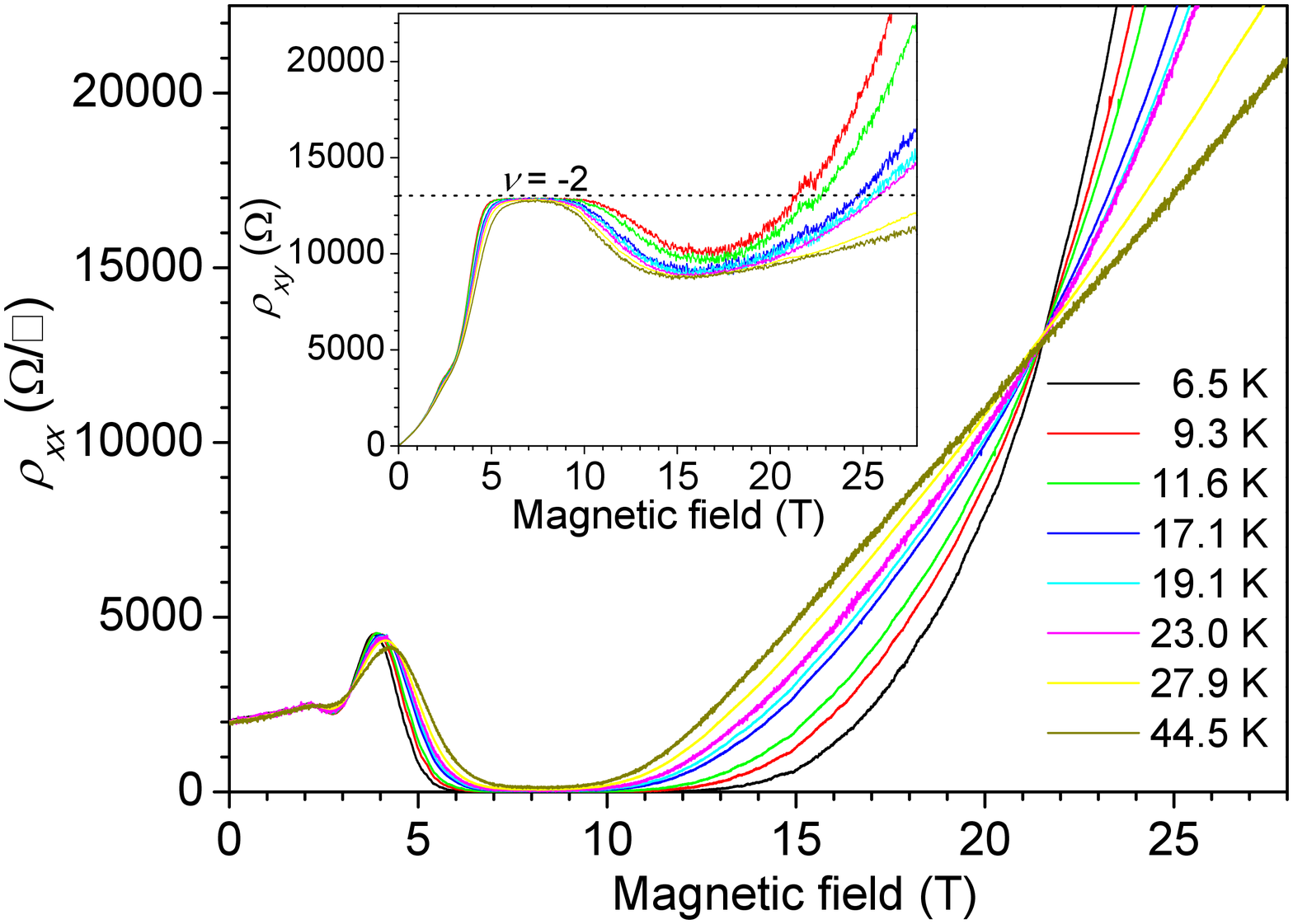}}
\caption{Longitudinal resistivity $\rho_{xx}$ as a function of the magnetic field for different temperatures at $V_{G} = 0$~V. A clear temperature-independent crossing point, separating the metallic and the insulating regions,  is present at $B_c = 21.56$~T with $\rho_{xx} = 0.504(4)$ R$_{H}$, denoting the plateau-insulator transition. The inset shows the Hall resistivity $\rho_{xy}$ measured in the temperature range at which the plateau $\nu = -2$ is resolved.}\label{resist0V}
\end{figure}

Fig.~\ref{resist0V} shows the longitudinal resistivity isotherms in the range of temperatures from $6.5$ to $44.5$~K with $V_G = 0$~V , $i.e.$ at $~4$~V from the Dirac point, while in the inset we report the Hall resistivity. At this $V_G$ the $n$ was found to be $6.8\times10^{11} cm^{-2}$ while $\mu$ was $ 4900$ $ cm^2V^{-1} s^{-1} $ at $6.5$~K. We note that in $\rho_{xx} $ we resolve the last plateau at $\nu = -2$ and we can observe the transition from the last Hall plateau to an insulating phase, with the typical T-independent crossing point (being the critical field $B_c = 21.56$~T). At this value of the magnetic field the longitudinal resistivity assumes the value $\rho_{xx} = 0.504(4)$ R$_{H}$ (being R$_{H} = 25.812$ $K\Omega$ the quantized Hall resistance in a semiconductor 2DEG) i.e. very close to the expected value taking into account the corrective factor $1/2$ due to the valley degeneracy characteristic of graphene~\cite{gusynin1}. This value is in very good agreement with the one found by Zhang \emph{et al.}~\cite{Zhang-MI} where it is clearly established that the PI takes place in the regime of the dissipative transport region that would correspond to magnetic fields where $R_{xx}>0.5$ R$_{H}$ and $\rho_{xx}\simeq0.5$. We have observed that at this gate voltage $R_{xx}>0.5$ R$_{H}$ occurs for magnetic fields greater than $16.6\,$T where $\rho_{xy}$ exhibits a broad minimum which marks the limit between the drop from the $\nu = -2$ QH plateau and the dissipative transport region in concordance with the previous work of Zhang \emph{et al.}~\cite{Zhang-MI}.

For magnetic fields smaller than $B_c$ the resistivity increases increasing T (metallic phase) while for magnetic fields greater than $B_c$ the resistivity decreases increasing T (insulating phase), thus revealing a PI transition. We can see that when $\rho_{xy}$ leaves the $\nu = -2$ at first decreases (above $B \sim 10 $ T) and then increases as observed and described by Poumirol \emph{et al.}~\cite{Escoffier}. At the same time $\rho_{xy}$ increases rapidly and diverges, a behavior that can be ascribed to a small misalignment of the Hall contacts, so that at high $B$ $\rho_{xx}$ dominates the Hall resistivity, due to the admixture of the two components. Despite $\rho_{xx}$ clearly diverges at high $B$ (a behavior more pronounced at low temperatures) we could not observe any trace of a Kosterlitz-Thouless (KT) transition,~\cite{KT} which actually is expected to occur only if the gate voltage is set exactly at the CNP.~\cite{Ong1}

\begin{figure}[ht]
\centerline{\includegraphics[width=90mm,clip]{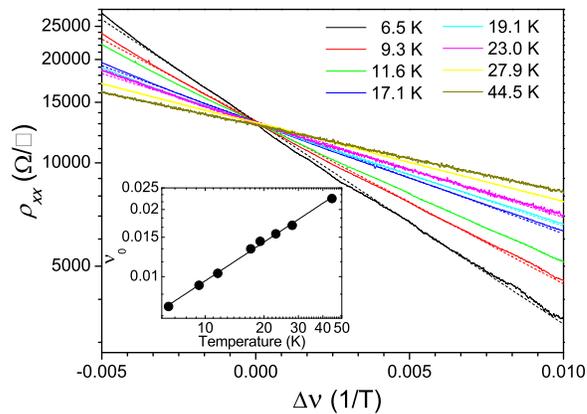}}
\caption{Longitudinal resistivity $\rho_{xx}$, in logarithmic scale, measured at different temperatures with $V_{G} = 0$~V as a function of $\Delta \nu$ (solid lines), and its best fit to~\ref{scaling1} (dashed lines). The inset shows, in double-log scale, $\nu_0$ as a function of temperature and its linear fit to~\ref{scaling-PI} (continuous line), which provides the critical exponent $\kappa = 0.58(1)$.}\label{fit0V}
\end{figure}

In order to obtain the critical exponent associated to the PI transition, for each temperature we renormalized the magnetic field using the expression $\Delta\nu = 1/B-1/B_c$ , and we extracted the value of $\nu_0$ from the fit of $\rho_{xx}$ to \ref{scaling1}. Figure~\ref{fit0V} reports the experimental $\rho_{xx}$ as a function of $\Delta\nu$ (continuous lines) at different temperatures, together with the best fits to \ref{scaling-PI} (dashed lines). The inset of Fig.~\ref{fit0V} reports the obtained values of $\nu_0$ (full dots) at the different temperatures, these values are reported without error bar since experimental noise was negligible. The fit to \ref{scaling-PI} (continuous line) allowed us to obtain the coefficient $\kappa = 0.58(1)$ , a value in very good agreement with the one that we obtained in a previous work~\cite{Amado-PI} where we measured with the same distance between the CNP and $V_G $ but in a more limited temperature range ($1.4$ - $4.2$~K). This behavior indicates that, when the system is sufficiently far away from the CNP, it behaves like a semiconductor 2DEG with alloy disorder.

\begin{figure}[ht]
\centerline{\includegraphics[width=90mm,clip]{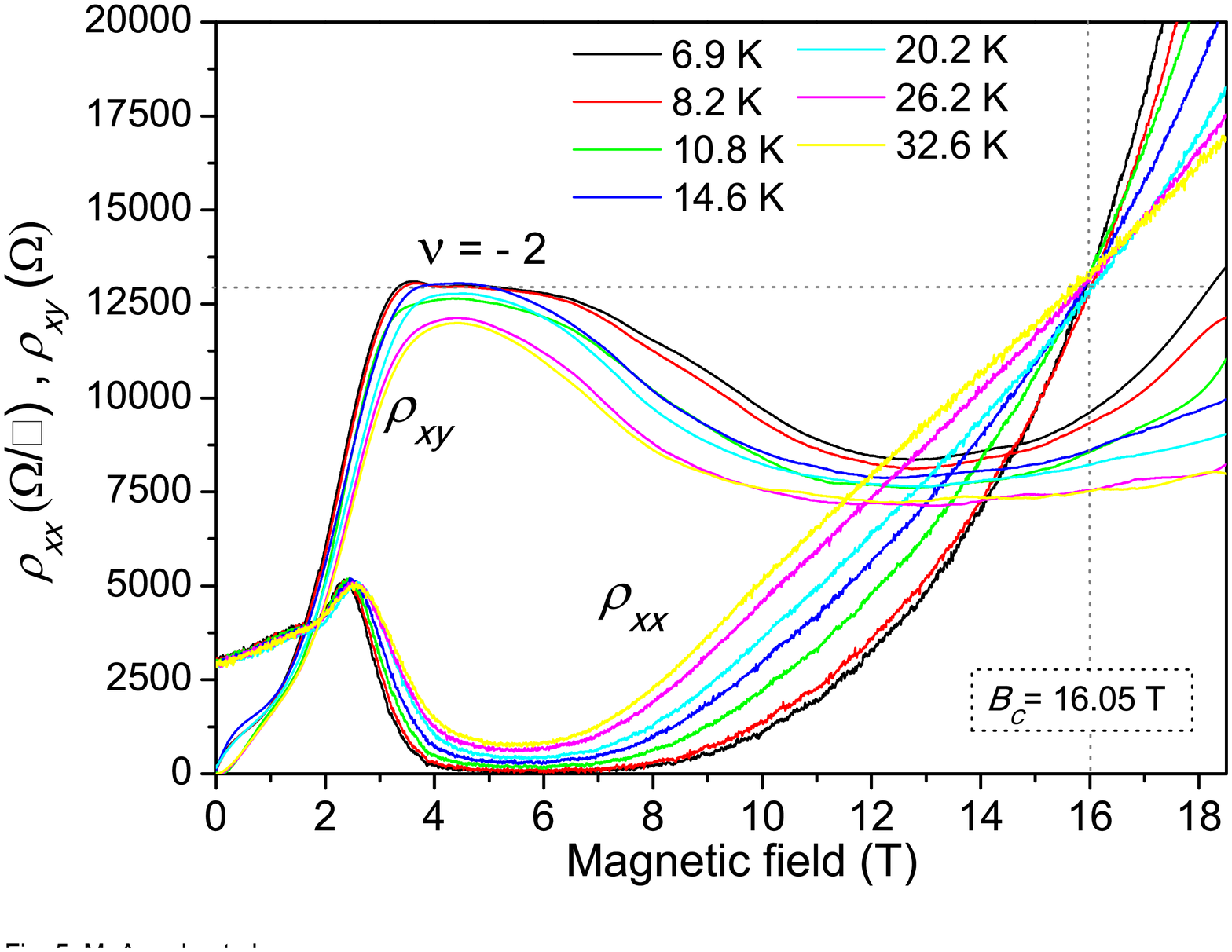}}
\caption{Longitudinal and Hall resistances isotherms as functions of the magnetic field, in the temperature range 6.9-32.6 K at $V_{G} = 2$~V. A temperature independent crossing point, which ideally separates the metallic and insulating phases, is present at $B_c = 16.05$~T with $\rho_{xx} = 0.51(1) $ R$_{H}$.}\label{resist2V}
\end{figure}

In figure~\ref{resist2V} we report the longitudinal and Hall resistivities measured in the temperature range $6.9$ - $32.6\,$ K, at $V_G = 2\,$V , with $n = 4\times10^{11} cm^{-2}$ and $\mu = 5400$ $cm^2V^{-1} s^{-1}$. We can observe a well resolved quantum Hall plateau at $\nu = -2$  at temperatures below $32.6$~ K and the PI transition at $\rho_{xx} = 0.51(1)$ R$_{H}$ (again, in good agreement with~\cite{Zhang-MI}) with the value of the critical magnetic field $B_c = 16.05$~T. We can also see that, also in this case, $\rho_{xy}$ decreases with $B$ after the $\nu = -2$ plateau showing a wide minimum at $B\sim$ 13 T where $R_{xx}$ enters into the dissipative transport regime~\cite{Zhang-MI}.

\begin{figure}[ht]
\centerline{\includegraphics[width=90mm,clip]{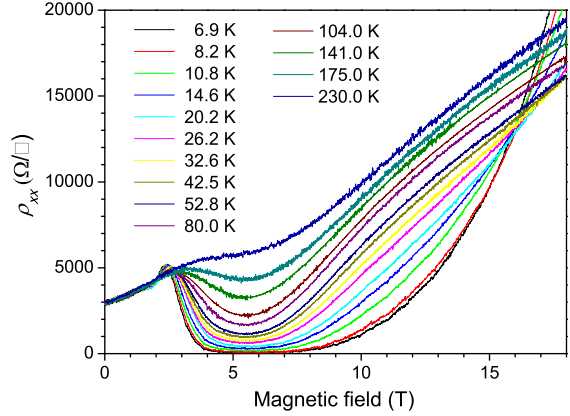}}
\caption{Longitudinal resistivity $\rho_{xx}$ as a function of the magnetic field measured at several temperatures at $V_{G} = 2$~V. The temperature independent crossing point disappears for  $B>32.6$~K, indicating the existence of a deactivation temperature of the plateau-insulator transition, which depends on the doping level of the sample.}\label{resist2V-range}
\end{figure}

In figure~\ref{resist2V-range} we show that for temperatures above $32.6\,$ K $\rho_{xx}$ did not take the value$0.5$ R$_{H}$ at B$_c$ but higher values and we affirm that for temperatures greater than $32.6\,$ K the quantum transport faded out and became classical thermal-activated thus disappearing the PI transition as has been recently observed also by Zhang \emph{et al.}~\cite{Zhang-MI}. We also found that at $V_G = 0$~V the transition can be  observed at temperatures up to $44.5$ K, so that this QPT in graphene is much more stable in than in semiconductor 2DEG where PI transition have been observed at temperatures up to $\sim 7$ K ~\cite{deLang}.  Moreover, our measurements did not provide indication of an onset of KT transition, suggesting that such a transition is dominant only when $V_G$ is set exactly at the CNP.

\begin{figure}[ht]
\centerline{\includegraphics[width=90mm,clip]{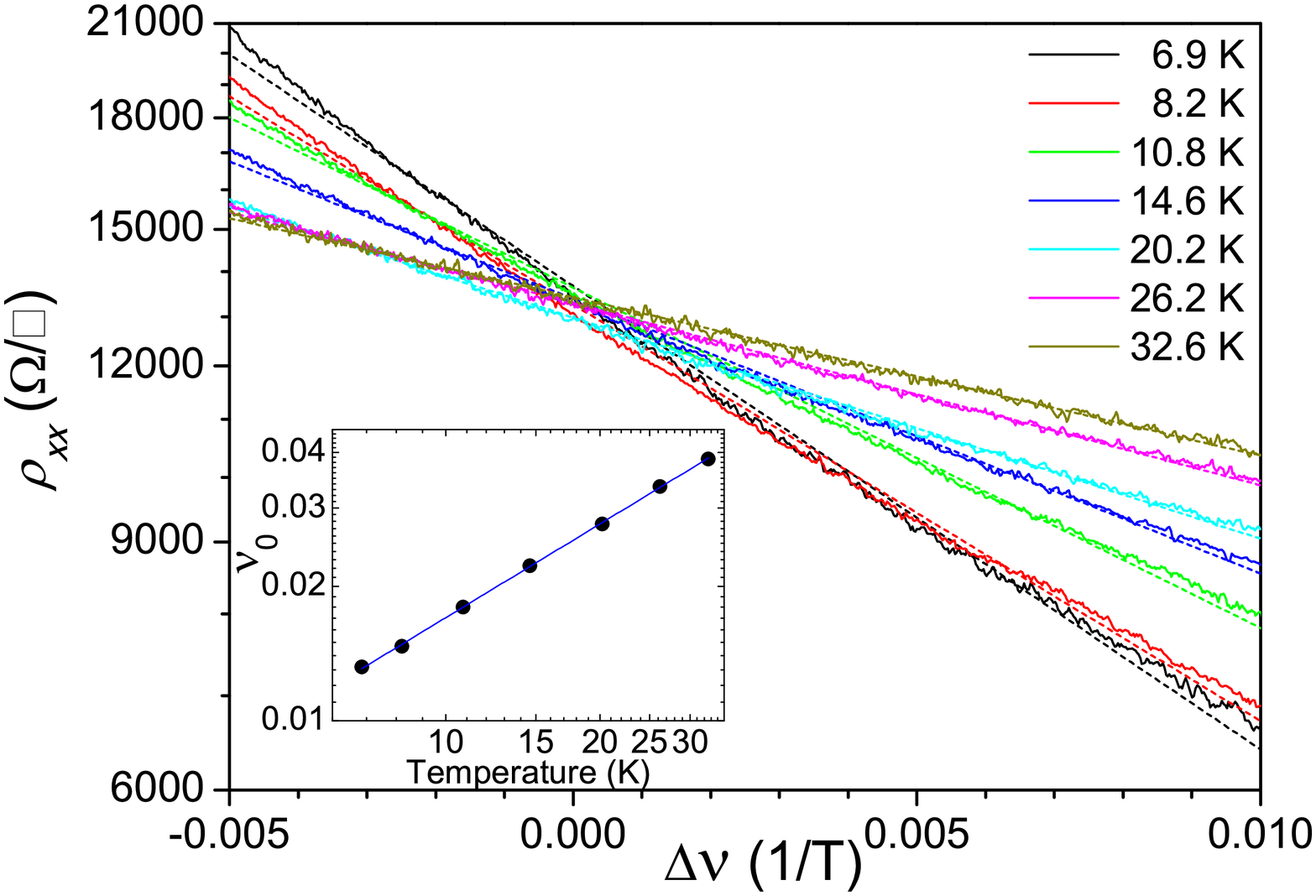}}
\caption{Longitudinal resistivity $\rho_{xx}$, in logarithmic scale, as a function of $\Delta \nu$ (dashed lines) at different temperatures and at $V_{G} = 2$~V. The solid lines are its best fit to \ref{scaling1}, which provide the values of $\nu_0$. The inset shows, in double-log scale, $\nu_0$ as a function of $T$ and its linear fit to \ref{scaling-PI} which provides the critical exponent $\kappa = 0.697(5)$.}\label{fit2V}
\end{figure}

Figure~\ref{fit2V} shows the $\rho_{xx}$ (solid lines) measured at different temperatures at $V_G = 2$~V together with their best fit to \ref{scaling1} (dashed lines), which give the values of $\nu_0$ that are reported (as full dots) in the inset of figure~\ref{fit2V}. The trend of $\nu_0$ with temperature was fitted to \ref{scaling-PI} giving the values $\kappa = 0.697(5)$.   This is different from the value $0.58$ obtained in the same sample but with higher carrier density. As we have previously pointed out, the contribution to the longitudinal resistivity in doped graphene is $\rho_L\propto1/n$ for long-range disorder.  When this contribution is comparable with or even larger than short-range disorder, the quantum localization crosses over toward the classical percolation. Our value is remarkably close to the values $\kappa=0.7$ of the classical full-percolation limit for the exponent, showing  that  for low doping level the scattering would be dominated by the \emph{e-e} interaction rather than by the charged impurities, which are instead expected to dominate at higher doping.~\cite{guinea-review}.

\section{Results and discussion: Plateau-Plateau transition}

\begin{figure}[ht]
\centerline{\includegraphics[width=90mm,clip]{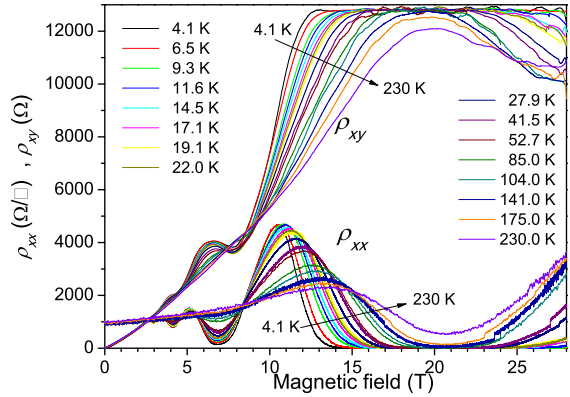}}
\caption{$\rho_{xx}$ and $\rho_{xy}$ resistivities as functions of the magnetic field at $V_{G}=-8$~V and at temperatures in the range 4.1-230 K.}\label{PP}
\end{figure}

In order to study the PP transition we tuned the gate voltage away from the CNP, namely at $V_G = -8$~V , where the density and Drude mobility were n$ = 1.8\times10^{12} cm^{-2}$ and $\mu = 3800$ $cm^2V^{-1} s^{-1}$, respectively. This value of  $V_G$ allowed us to observe the Shubnikov-de Haas peaks of $\rho_{xx}$ and the quantum Hall plateaus at $\nu = -10$, $-6$ and $-2$ of $\rho_{xy}$ in the temperature range $4.1-230$~K that we show in figure~\ref{PP}. The PP transition remained clearly observable almost up to the maximum temperature reached ($230$~K), in agreement with the finding of Giesbers \emph{et al.}~\cite{ppgiesbers}. In our experiments we followed  different procedure with respect to the one of Ref.~\cite{ppgiesbers}, since we used the magnetic field, instead of  $V_G $, as driving parameter. The parameter $\kappa$ has been calculated using \ref{power-law-K-B}, from the temperature dependence of the maximum slope in the derivative  $\left( \frac{\partial \rho_{xy}}{\partial B}\right)_{max}$ for the transitions $\nu = -10 \rightarrow -6$ and $\nu = -6 \rightarrow -2$ and from the temperature dependence of the full width at half maximum (FWHM) of the SdH peak ($\Delta B$) for the transition $\nu = -6 \rightarrow -2$.

\begin{figure}[ht]
\centerline{\includegraphics[width=120mm,clip]{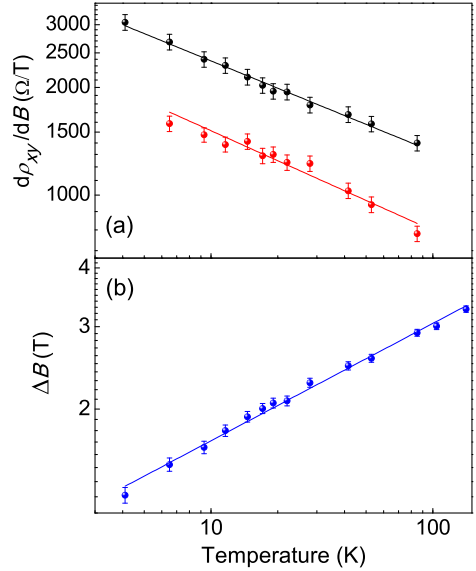}}
\caption{Values for $\left( \frac{\partial \rho_{xy}}{\partial B}\right)_{max}$ $a$ and  $\Delta B$ $b$ versus temperature in double-log scale. In $a$ we report on the data for the PP transitions $\nu=-10 \rightarrow -6$ (red full dots) and $\nu=-6 \rightarrow -2$ (black full dots). The solid lined are the corresponding least squares fits to~(\ref{power-law-K-rho}). In $b$ we report on the $\Delta B$ for the transition $\nu=-6 \rightarrow -2$ (blue full dots) together with the least squares fit to~(\ref{power-law-K-B}) (blue solid line).}\label{fitPP}
\end{figure}

Figure~\ref{fitPP}$a$ shows $\left( \frac{\partial \rho_{xy}}{\partial B}\right)_{max}$, for the transitions $\nu=-6 \rightarrow -2$ (black full dots) and for the transition$\nu=-10 \rightarrow -6$ (red full dots)  together with their least squares fits to~(\ref{power-law-K-rho}) (black and red lines). The $\kappa$ values obtained from this fits are $\kappa=0.254(6)$ for the $\nu=-6 \rightarrow -2$ transition, and $\kappa=0.28(2)$, for the $\nu=-10 \rightarrow -6$ one. Both results differ considerably from the typical value of the Anderson transition $0.42$ found in graphene by Giesbers \emph{et al.}~\cite{ppgiesbers}, indicating a non-universality of the PP critical exponent in such material. We believe that this is due to the effect of the different type of disorder on the charge carrier transport in graphene. In this frame, graphene would behave as a semiconductor 2DEG according to the Anderson model in presence of short range disorder, while it behaves as a semiconductor 2DEG following a percolation scenario in presence of long range disorder. Therefore the study of the critical exponent of the quantum phase transitions of the magneto-transport in graphene, allows to determine type of disorder that control the charier transport.

Figure~\ref{fitPP}$b$ shows the value of $\Delta B$ as a function of temperature for the transition $\nu=-6 \rightarrow -2$  (blue full dots). In this case $\Delta B$ as been taken as twice the distance between the maximum of the SdH peak and the and its right half maximum, in order to avoid its slight asymmetry.  The best squares fit to~(\ref{power-law-K-B}) (blue straight line) gives $\kappa=0.25(1)$, in perfect agreement with the one obtained from the previous the analysis of $\rho_{xy}$.

\section{Conclusions}

Summarizing, we have studied the magneto-transport of graphene and the nature of the quantum phase transitions that characterize the quantum Hall regime. Concerning the $\nu=-2$ to $\nu=0$ plateau-insulator transition, we have observed the QPT up to $44.5$~K, pointing out its robustness. Using the standard scaling theory analysis we have experimentally obtained its critical exponent, finding that it is consistent with the universal value of semiconductor 2DEG when the sample is doped away from the Dirac point $\kappa=0.58(1)$, while it tends to the classical full percolation limit value $\kappa=0.697(5)$ when $V_G$ approaches the charge neutrality point. Our measurement indicates that the electron-electron interaction plays a major role on the magneto-transport close to the charge neutrality point, modifying the value of the critical exponent. Moreover we have found no traces of Kosterlitz-Thouless transitions approaching the CNP in contrast with the previous finding~\cite{Ong1}. As regards the other QPT, the plateau-plateau one, we have studied it up to $230$~K obtaining the value for the critical exponent $\kappa=0.25$, and we revealed its non universality. Our results evidence that magneto-transport measurements in graphene in the quantum Hall regime, and the analysis of the observed quantum phase transitions, are powerful tools to elucidate the interplay between disorder and the electron-electron interaction, which controls the charge transport in the graphene nano-devices.

\section{Acknowledgements}

We acknowledge fruitful discussions with Alberto Rodr\'{i}guez, Paco Guinea and Luis Brey. This work has been supported by the following projects: Cariplo Foundation QUANTDEV, Ministerio de Ciencia e Innovaci\'on FIS2009-07880 and PCT310000-2009-3, Junta de Castilla y Le\'on SA049A10, and European Union CTA-228043-EuroMagNET II Programme.


\section*{References}
\begin{thebibliography}{10}
\bibitem{weiprb} Wei H P, Tsui D C and Pruisken A M M 1986 \PR \emph{B} {\bf 33} 1488

\bibitem{tsui} Wei H P, Tsui D C, Paalanen M A and Pruisken A M M 1988 \PRL {\bf 61} 1294

\bibitem{pruisken-pp}Pruisken A M M 1988 \PRL {\bf 61} 1297

\bibitem{CC} Chalker J T and Coddington P D 1988 \JPC {\bf21} 2665

\bibitem{Lee} Lee D-H, Wang Z and Kivelson S 1993 \PRL {\bf 70} 4130

\bibitem{Cho} Cho S and Fisher M P A 1997 \PR \emph{B} {\bf 55} 1025

\bibitem{Slevin} Slevin K and Ohtsuki T, 2009 \PR \emph{B} {\bf 80} 041304

\bibitem{Amado-CC} Amado M, Malyshev A V, Sedrakyan A and Dom\'{i}nguez-Adame F 2011 \PRL {\bf 107} 066402

\bibitem{Anissimova} Anissimova S, Kravchenko S V, Punnoose A,Finkelstein A M and Klapwijk T M. 2007 \emph{Nature Physics}  {\bf 3} 707

\bibitem{Martin} Martin J, Akerman N,Ulbricht G, Lohmann T, Klitzing von K,Smet J H and Yacoby A 2009 \emph{Nature Physics}  {\bf 5} 669

\bibitem{Lee2} Lee  D-H and Wang Z, 1996 \PRL {\bf 76}   4014


\bibitem{Baranov} Baranov M A,  Burmistrov I  S and Pruisken A M M, 2002 \PR \emph{B} {\bf 66} 075317

\bibitem{Huckestein} Huckestein B 1995 \RMP {\bf 67} 357

\bibitem{Wanli} Li WanLi, Vicente C L, Xia J S, Pan W, Tsui D C, Pfeiffer L N and West K W 2009 \PRL \textbf{102} 216801

\bibitem{Wanlialloy} Li Wanli, Cs\'athy G A, Tsui D C, Pfeiffer L N and West K W 2005 \PRL {\bf 94} 206807

\bibitem{wakabayashipp} Wakabayashi J, Kwon H C and Park J C 1994 \SSC {\bf 98} 821

\bibitem{koch} Koch S, Haug R J, von Klitzing K and Ploog K 1991 \PR \emph{B} {\bf 43} 6828

\bibitem{Hilke1} Hilke M, Shahar D, Song S H, Tsui D C, Xie Y H and Monroe D 1998 \emph{Nature} {\bf 395} 675

\bibitem{Schaijk} van Schaijk R T F, de Visser A, Olsthoorn S M, Wei H P and Pruisken A M M 2000 \PRL {\bf 84} 1567

\bibitem{deLang} de Lang D T N, Ponomarenko L A, de Visser A, Possanzini C, Olsthoorn S M and Pruisken A M M, 2002 \emph{Physica E} {\bf 12} 666

\bibitem{ponomarenko1} Ponomarenko L A, de Lang D T N, de Visser A, Maude D K, Possanzini C, Olsthoorn S M and Pruisken A M M 2004 \emph{Physica E} {\bf 22} 236

\bibitem{ponomarenkoTH} Ponomarenko  L A "Experimental Aspects of Quantum Criticality in the Quantum Hall regime" 2005 Ph.D. Thesis, Universiteit van Amsterdam, ISBN 90-5776-144-0

\bibitem{Pruisken} Pruisken A M M, de Lang D T N, Ponomarenko L A and de Visser A 2006 \SSC {\bf 137} 540

\bibitem{novoselov2005} Novoselov K S, Geim A K, Morozov S V, Jiang D, Katsnelson M I, Grigorieva I V, Dubonos S V and Firsov A A 2005 \emph{Nature} {\bf438} 197

\bibitem{stormer2005} Zhang Y, Tan Y-W, Stormer H L and Kim P 2005 \emph{Nature} {\bf438} 201

\bibitem{guinea2} Peres N M R, Guinea F and Castro Neto A H 2006 \PR \emph{B} {\bf 73} 125411

\bibitem{gusynin1} Gusynin V P and Sharapov S G 2004 \PRL {\bf 95} 146801

\bibitem{Zhang} Zhang Y, Jiang Z, Small J P, Purewal M S, Tan Y-W, Fazlollahi M, Chudow J D, Jaszczak J A, Stormer H L and Kim P 2006 \PRL {\bf 96} 136806

\bibitem{Abanin} Abanin D A, Novoselov K S, Zeitler U, Lee P A, Geim A K and Levitov L S 2007 \PRL {\bf 98} 196806

\bibitem{Jiang} Jiang Z, Zhang Y, Stormer H L and Kim P 2007 \PRL {\bf 99} 106802

\bibitem{Ong1} Checkelsky J G, Li L and Ong N P 2008 \PRL {\bf 100} 206801; {\em ibid} 2009 \PR \emph{B} {\bf 79} 115434

\bibitem{Giesbers} Giesbers A J M, Ponomarenko L A, Novoselov K S, Geim A K, Katsnelson M I, Maan J C and Zeitler U 2009 \PR \emph{B} {\bf 80} 201403(R)

\bibitem{Du} Du X, Skachko I, Duerr F, Luican A and Andrei E Y 2009 \emph{Nature} {\bf 462} 192

\bibitem{Mucciolo} Mucciolo E R and Lewenkopf C H 2010 \JPC {\bf 22} 273201

\bibitem{Wiedmann} Wiedmann S, van Elferen H J, Kurganova E V, Katsnelson M I, Giesbers A J M, eligura A, van Wees B J, Gorbachev R V, Novoselov K S, Maan J C and Zeitler U 2001 \PR \emph{B} \textbf{84} 115314

\bibitem{Poumirol} Poumirol J-M, Escoffier W, Kumar A, Goiran M, Raquet B and Broto J M 2010 \NJP \textbf{12} 083006

\bibitem{Zaliznyak} Zhang L, Camacho J, Cao H, Chen Y P, Khodas M, Kharzeev D E, Tsvelik A M, Valla T and Zaliznyak I A 2009 \PR \emph{B} \textbf{80} 241412(R)

\bibitem{Amado-PI} Amado M, Diez E, L\'opez-Romero D, Rossella F, Caridad J M, Dionigi F, Bellani V and Maude D K 2010 \NJP \textbf{12} 053004
\bibitem{Morozov} Morozov S M, Novoselov K S, Katsnelson M I, Schedin F, Elias D C, Jaszczak J A and Geim A K, Phys. Rev. Lett. \textbf{100}, 016602 (2008).

\bibitem{ppgiesbers} Giesbers A J M, Zeitler U, Ponomarenko L A, Yang R, Novoselov K S, Geim A K and Maan J C 2009 \PR \emph{B} \textbf{80} 241411(R)

\bibitem{Shahar1} Shahar D, Hilke M, Li C C, Tsui D C, Sondhi S L, Cunningham J E and Razeghi M 1998 \SSC{\bf 107} 19

\bibitem{Escoffier} Poumirol J-M, Escoffier W, Kumar A, Raquet B and Goiran M 2010 \PR \emph{B} \textbf{82} 121401(R)

\bibitem{Zhang-MI} Zhang L, Zhang Y, Khodas M, Valla T and Zaliznyak I A 2010 \PRL {\bf 105} 046804

\bibitem{KT} Kosterlitz J M and Thouless D J 1973 \JPC {\bf 6} 1181

\bibitem{guinea-review} Kotov V N, Uchoa B, Pereira V M, Castro Neto A H and Guinea F arXiv:1012.3484v.1
\end{thebibliography}
\end{document}